\documentstyle[multicol,aps,epsf]{revtex}

\begin{document}

\title{Scattering Theory of Photon-Assisted Electron Transport}

\author{Morten Holm Pedersen and Markus B\"uttiker}
\address{D\'epartement de Physique Th\'eorique,
Universit\'e de Gen\`eve, 1211 Gen\`eve 4, Switzerland}

\date{\today}
\maketitle

\begin{abstract}
The scattering matrix approach to phase-coherent transport is
generalized to nonlinear ac-transport. 
In photon-assisted electron transport it is often only the dc-component 
of the current that is of experimental interest. 
But ac-currents at all frequencies exist independently of 
whether they are measured or not. We present a theory of 
photon-assisted electron transport which is charge and current 
conserving for all Fourier components of the current. 
We find that the photo-current can be considered as an up- and down-conversion
of the harmonic potentials associated with the displacement 
currents. 
As an example explicit calculations are presented for a resonant double barrier  
coupled to two reservoirs and capacitively coupled to a gate. Two experimental
situations are considered: in the first case the ac-field is applied via a gate,
and in the second case one of the contact potentials is modulated. 
For the first case we 
show that the relative weight of the conduction sidebands varies with
the screening properties of the system. 
In contrast to the non-interacting case the relative weights
are not determined by Bessel functions. 
Moreover, interactions can give rise 
to an asymmetry between absorption and
emission peaks.
In the contact driven case, the theory predicts a zero-bias current 
proportional to the asymmetry of the double barrier. This is in contrast  
to the discussion of Tien and Gordon which, in violation of basic symmetry principles, 
predicts a zero-bias current also for a symmetric double barrier.
% 73.23.-b: Mesoscopic systems
% 85.30.Vw: Low dimensional quantum devices
% 73.40.Gk: Tunneling
% 72.10.-d: Theory of electronic transport; scattering mechanisms
\end{abstract}

\begin{multicols}{2}

\section{Introduction}

Photon-assisted tunneling has been of interest 
since the work of Tien and Gordon \cite{tien} and 
Tucker \cite{tucker}. Carrier transmission 
through barriers with oscillating potentials has 
been analyzed to find the traversal time for tunneling \cite{buttiker82}.
Recently photon-assisted tunneling has found 
renewed interest in the field of mesoscopic physics
stimulated by theoretical work by Bruder and Schoeller \cite{bruder}
and experiments on quantum dots by Kouwenhoven and McEuen et al. \cite{kouwenhoven},
and by experiments on superlattices by the group of 
Allen et al. \cite{guimaraes,drexler,keay}.
Typically of interest 
\cite{bruder,kouwenhoven,drexler,keay,datta,jauho94,wacker,wagner,wagner2,aguado,aguado2,avishai,flensberg,zhao,sun,martin,stafford,verghese,stoof,gorelik}
is the zero-frequency current component induced in response to an 
oscillating voltage. 
Theoretical treatments of
photon-assisted electron transport often assume that the driving field is 
known and equals
the external field. However, the long-range Coulomb interaction
will screen the external field and generates an internal potential that can be quite
different from the applied potential.
Similarly, since it is the dc-component
which is measured,
one might think that displacement currents 
play no role. However, the 
dc-component is a consequence of nonlinearities in the conduction process. 
Clearly, in such a conductor, the current has not only a dc-component, 
but also currents at the frequency of the oscillating voltage and its higher
harmonics. Not only are the dc-currents conserved but also 
the currents at the oscillation frequency and at its higher 
harmonics. Consequently a theory is needed 
in which all frequency components of the current 
are treated self-consistently. 
Such a theory is developed below. It leads to the conclusion 
that the photo-current
is induced by a non-linear up- and down-conversion of the electric-fields (potentials) 
associated 
with the displacement current. 
Bessel functions are often a hall mark \cite{tien} 
of the discussion of photon-assisted 
tunneling: However, in the self-consistent theory 
discussed below Bessel functions cannot in general describe the relative weights
of the sideband peaks since the argument of the Bessel functions is not 
invariant
under an overall potential shift. 
Since in nonstationary conditions
charge accumulation occurs and causes induced fields, 
a selfconsistent treatment
of the electron-electron interactions is important. 
The issues are similar for theories and 
experiments which investigate photon-assisted process not in the dc-current 
but in its fluctuations \cite{lesovik,schoelkopf}. Here we emphasize mainly
the average current and address the fluctuation spectra only briefly
in an appendix.

A convenient description of conduction processes in mesoscopic systems which
incorporates the role of contacts and permits to investigate directly 
the phase-coherent transmission from one reservoir to another, is the
scattering matrix approach \cite{landauer,buttiker86}. 
The description of linear ac-conduction in response to oscillating potentials 
and consideration of the long range Coulomb interaction has already been 
discussed both for the case of zero-dimensional systems \cite{buttiker93,pretre} 
and for extended systems for which one needs to discuss
the entire potential landscape \cite{buttiker932,blanter98}.  
A review of this subject can be found in Ref.~\cite{buttiker96}.
Here we generalize the scattering matrix approach to take into account the 
nonlinear dependence on oscillating
potentials. First we consider the response of the electrons to a
potential applied only to the contacts of the sample,
assuming the internal potential is kept fixed. 
The response to
the total potential will, in a subsequent step,
be calculated selfconsistently in random-phase-approximation (RPA).
The resulting charge and current conserving theory will be used to investigate the
photo-induced dc-current in a resonant tunneling barrier. As function of 
Fermi
energy and frequency we find large differences between the induced internal potential
and the external applied potential, showing that long range Coulomb forces are
important for photon-assisted tunneling in mesoscopic systems.
Furthermore, interactions can give rise to an asymmetry between absorption and
emission peaks, as well as changing the distance between peaks from a multiple of
photon quanta to a distance depending on screening properties.

Our discussion is complementary to works which model 
interactions based on a Hamiltonian suitable to describe Coulomb blockade effects. 
The work of Bruder and Schoeller \cite{bruder} also considers coupling to a gate and 
also considers displacement currents. In principle, all the questions addressed here 
can be investigated within such a framework. The scattering approach used here has the 
advantage  that it is not limited to the tunneling regime but can also be applied to 
conductors which are strongly coupled to reservoirs (ballistic
or metallic diffusive wires, etc). The RPA treatment as it is formulated 
below does have the disadvantage that it is not an appropriate description in 
the case when charge quantization effects (Coulomb blockade) are important. 
However, its conceptual clarity makes the RPA treatment a useful point of 
reference for comparision of different theoretical discussions. 

The basic view taken here is the same as that used for the discussion of
dc-conductances\cite{buttiker86} and ac-conductance \cite{buttiker93}. What is needed is the
connection between currents at the contacts of the structure and the
voltages at these contacts. Either the currents or the voltages can be
controlled. As in the discussion of the ac-conductance it is necessary
to consider not only the mesoscopic conductor itself but all nearby metallic
bodies (gates and capacitors) which interact via long range Coulomb forces
with the mesoscopic conductor. Let $\alpha$ label all the relevant contacts.
The current at contact $\alpha$ can be written in terms of its Fourier
components $I_\alpha(n\omega)$. 
Here $n = 0$ is the dc component of the currents, 
and $n = \pm 1$ are the 
Fourier components at the driving 
frequency. Nonlinearities lead to higher harmonics 
$n = \pm 2, 3, ..$.  Similarly, the voltage at contact $\alpha$
has the Fourier components $V_\alpha(n\omega)$. We emphasize that the voltage
of a contact is only a well defined quantity if local electric fields deep
inside the contact vanish. There must, therefore, exist a Gauss volume which
encloses the mesoscopic conductor \cite{buttiker93}. The electric flux 
through this Gauss volume
vanishes. As a consequence the total charge $Q$ inside the volume is 
conserved \cite{buttiker93}.
Charge conservation, and current conservation, apply to each
Fourier component separately.
In particular, we must have that the 
total charge within the Gauss volume 
vanishes at each frequency, 
\begin{eqnarray}
Q_\alpha(n\omega)=0.
\label{fundamental}
\end{eqnarray}
A theory for which this holds gives currents which depend
ultimately only on voltage differences. We call such a theory of electric
conductance gauge invariant 
\cite{Christen97}. To be definite let us introduce an expansion
parameter $\epsilon$.  
We take the 
Fourier components of the first harmonic $V_\alpha(\omega)$
proportional to $\epsilon$ and expand 
the currents in powers of $\epsilon$.  The second harmonic voltages $V_\alpha(2\omega)$
describing two-photon processes are then proportional to $\epsilon^2$. Below we
write the relationship between currents and voltages up to second order in 
$\epsilon$. The expansion coefficients are conductances 
$g_{\alpha\beta\gamma}(n\omega,m\omega)$ which give the current at contact
$\alpha$ in response to a voltage $V_\beta(n\omega)$ at contact $\beta$ at a
frequency $n\omega$ and a voltage a contact $\gamma$ at a frequency $m\omega$.
The overall dc-current is
\begin{eqnarray}
    I_\alpha(0) &=& I^{dc}_\alpha[\{V_\beta(0)\}] + 
	I^{ph}_\alpha[0;\{V_\beta(0)\}], \label{w0} \\
    I^{ph}_\alpha[0;\{V_\beta(0)\}] &=&
	\sum_{\beta\gamma} g_{\alpha\beta\gamma} [\omega,-\omega;\{V_\delta(0)\}]
	V_\beta(\omega) V_\gamma^*(\omega). 
\end{eqnarray}
The first term of Eq.~(\ref{w0}), 
$I^{dc}_\alpha[\{V_\beta(0)\}]$, is the direct current that would be measured 
in the presence of purely static voltages $V_\beta(0)$, $\beta = 1, 2,..$
applied to the different contacts of the sample. 
In the following, for the direct current, 
we retain the full dependence to all
orders in the static applied voltages.
If the dc-current $I^{dc}_\alpha[\{V_\beta(0)\}]$
is expanded in powers of the applied voltage then the 
terms linear in the applied voltages determine  
the dc-conductance matrix
$g_{\alpha\beta}(0)$ and the terms quadratic in the applied voltages 
are the dc-rectification conductances $g_{\alpha\beta\gamma}(0)$,
discussed by Christen and one of the authors \cite{christen96,buttiker93},
which determine
the leading order non-linearity of the dc I-V-characteristic \cite{note1}.  
In addition to these contributions to the dc-current 
which characterize the purely stationary transport
there is now also a contribution to the dc-current 
due to the photon-assisted processes, $I^{ph}_\alpha[0;\{V_\beta(0)\}]$.
In particular, to second order in the applied ac-voltages $V_{\beta}(\omega)$,
carriers which emit and re-absorb (virtual) photons 
are determined by the dc-photo-conductance 
$g_{\alpha\beta\gamma}[\omega,-\omega;\{V_\delta(0)\}]$
which depends in general also on the dc-voltages $V_\delta(0)$.
These conductance
coefficients represent an up- and down-conversion of the first harmonic voltages.

The current at the frequency of the oscillating potential
is in general composed both of a particle current and of a displacement 
current. To be brief we call this current simply the displacement current.
To linear order in our expansion parameter it is given by   
\begin{equation} \label{w1}
    I_\alpha(\omega) = \sum_\beta g_{\alpha\beta}[\omega;\{V_\gamma(0)\}]
	V_\beta(\omega).
\end{equation}
Here expanding $g_{\alpha\beta}[\omega;\{V_\gamma(0)\}]$ in the dc voltages yields
the equilibrium admittance \cite{pretre}
of the mesoscopic structure $g_{\alpha\beta}(\omega)$ and
the dc-ac-rectification conductance $g_{\alpha\beta\gamma}(\omega;0)$.

The current at $2\omega$ is 
\begin{eqnarray}
    I_\alpha(2\omega) &=& \sum_\beta g_{\alpha\beta}[2\omega;\{V_\gamma(0)\}] 
	V_\beta(2\omega) \nonumber \\
	&& +
	\sum_{\beta\gamma} g_{\alpha\beta\gamma}[\omega,\omega;\{V_\delta(0)\}]
	V_\beta(\omega) V_\gamma(\omega)  \label{w2}
\end{eqnarray}
determined by a second harmonic conductance 
$g_{\alpha\beta}[2\omega;\{V_\gamma(0)\}]$ and a
non-linear up-conversion conductance 
$g_{\alpha\beta\gamma}[\omega,\omega;\{V_\delta(0)\}]$
whereby a second harmonic current is generated due to a non-linear combination
of first harmonic voltages. We emphasize that the expansion given here can in
principle be carried further to an arbitrary order in $\epsilon$.
Our task is to find explicit expressions for the (non-linear) ac-conductances
defined in Eqs.~(\ref{w0} - \ref{w2}). It is useful, to state first 
a number of general properties of these conductances. 

Current conservation holds for each Fourier component separately. Furthermore,
since we can break off the expansion at any order, current conservation restricts
each type of conductance coefficient in Eqs.~(\ref{w0}-\ref{w2}). An additional
restriction imposed on these conductance coefficients arises due to the fact that
a voltage $V(n\omega)$ which is applied to all contacts simultaneously can not have
a physical effect. As a consequence the conductances obey the 
sum rules \cite{christen96,buttiker93}
\begin{equation}
    \sum_\alpha g_{\alpha\beta}(k\omega) = \sum_\beta g_{\alpha\beta}(j\omega) = 0
\end{equation}
for $k,j\in {\cal N}$. Similarly, the second order coefficients obey
\begin{eqnarray}
    \sum_\alpha g_{\alpha\beta\gamma}(k\omega,j\omega) &=& 
	\sum_\beta g_{\alpha\beta\gamma}(k\omega,j\omega) \nonumber \\
	&& = \sum_\gamma g_{\alpha\beta\gamma}(k\omega,j\omega) = 0.
\end{eqnarray}
These sum rules guarantee that the final result will depend on voltage differences
only.

\begin{figure}
\narrowtext
\epsfysize=6cm
\epsfxsize=7cm
\centerline{\epsffile{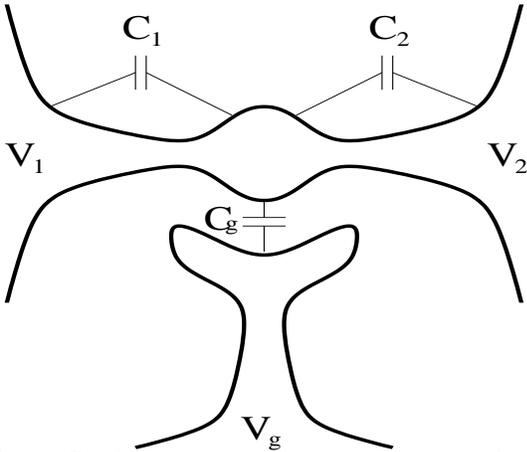}}
\caption{ \label{structure}
Conductor connected to two contacts and coupled capactively to a gate.}
\end{figure}

Eqs.~(\ref{w0} - \ref{w2}) are completely general and are applicable to any phase-coherent
multi-terminal conductor. We now discuss these general relations 
for the case of a two-terminal conductor capacitively coupled to 
a gate, a situation sketched in Fig.~\ref{structure}, in the limit $C_1=C_2=0$.
This simple arrangement permits us already to point 
to the connection between photo-currents and displacement currents. 
We are interested
in the photo-current generated by a sinusoidal oscillation of the voltage
$V_g(\omega)$ at the gate. First consider the displacement current. The oscillating
gate couples with the conductor in a purely capacitive manner. Therefore, the
$g_{\alpha g}[\omega;\{V_\beta(0)\}]$ 
describe capacitive currents and we can write
$g_{\alpha g}[\omega;\{V_\beta(0)\}]= -i\omega C_{\alpha g}[\omega;\{V_\beta(0)\}]$. 
We emphasize that this is a global
transport coefficient which connects the voltage at one contact to the current
at another contact. As a consequence, the capacitance coefficients are not of
a purely geometrical nature but can be strong functions of magnetic field
and the dc gate voltage \cite{buttiker93,Chen,Christen97,Moon}. Thus, the current at
contact $\alpha$ is determined by
\begin{equation}
    I_\alpha(\omega) = -i\omega C_{\alpha g}[\omega;\{V_\beta(0)\}]  
	V_g(\omega).
\end{equation}
Next, consider the dc-photo-current generated by this arrangement
\begin{equation}
   I^{ph}_\alpha(0) = g_{\alpha gg}[\omega,-\omega;\{V_\beta(0)\}] |V_g(\omega)|^2.
\end{equation}
With any of the equations for the displacement current we can eliminate the
oscillating gate voltage entirely and find
\begin{equation} \label{ph_current}
    I^{ph}_\alpha(0) = \frac{1}{\omega^2} 
	\frac{g_{\alpha gg}[\omega,-\omega;\{V_\beta(0)\}]}
	{ | C_{\alpha g}[\omega;\{V_\beta(0)\}]|^2 }
	|I_\alpha(\omega)|^2.
\end{equation}
Thus to second order
in the oscillating voltages, the photo-current is directly related to the displacement 
current.
Since the displacement current is not a property of a non-interacting system
but is in an essential way determined by the long range Coulomb interaction,
so similarly, the long range Coulomb interaction must play an essential role
in determining the photo-current. Note that the photo-conductance which enters 
Eq.~(\ref{ph_current}) is also proportional to $\omega^2$ and the photo-current 
given in Eq.~(\ref{ph_current}) therefore has a well defined zero-frequency limit. 

Now we proceed to find explicit expressions for the non-linear 
conductances introduced above.

\section{Oscillating contact potentials: External response}

We consider a conductor with voltages which oscillate in time applied to 
the contacts of the sample or to nearby capacitors. 
First we evaluate the response of non-interacting particles with the internal
potential kept fixed. Only the response to the total potential has physical
meaning, however, these results are needed in the next section for treating
the problem with interactions.

The current 
operator for current incident in contact $\alpha$ in a mesoscopic 
system can be written as \cite{buttiker92}
\begin{eqnarray}
    \hat{I}_\alpha(t) &=& \frac{e}{h} \int dE \int dE' \left[
	\hat{\bf a}_\alpha^\dagger(E) \hat{\bf a}_\alpha(E') \right. \nonumber \\
	&& - \left.
	\hat{\bf b}^\dagger_\alpha(E) \hat{\bf b}_\alpha(E') \right]
	e^{i\frac{E-E'}{\hbar} t} 
\label{current_op}
\end{eqnarray}
where $\hat{\bf a}_\alpha$ and $\hat{\bf b}_\alpha$ are vectors
of operators with components $\hat{a}_{\alpha n}$ and $\hat{b}_{\alpha n}$.
Here $\hat{a}_{\alpha n}$ annihilates an incoming carrier
in channel $n$ in lead $\alpha$ and  $\hat{b}_{\alpha n}$ annihilates 
an outgoing carrier in channel $n$ in lead $\alpha$.
Eq. (\ref{current_op}) applies for frequencies $(E-E')/\hbar$ small 
compared to the Fermi energy.

The incoming and outgoing waves are related by the scattering 
matrix \cite{buttiker92} ${\bf s}_{\alpha\beta}$ via, 
$\hat{\bf b}_\alpha=\sum_\beta {\bf s}_{\alpha\beta} \hat{\bf a}_\beta$.
In a multichannel conductor the s-matrix has dimensions 
$N_\alpha\times N_\beta$ for leads with $N_\alpha$ and $N_\beta$ channels.
Here, and in the following, greek indices run over all contacts of the conductors.

Let us now suppose that a potential variation is applied to reservoir 
$\alpha$. The potential is 
$eU_\alpha(t)=eV_\alpha(\omega) \cos\omega t$, where 
$V_\alpha(\omega)$ is the modulation amplitude.
With this potential the solution to the single-particle Schr\"odinger equation
at energy $E$ in $\alpha$ is 
\begin{equation}
    \psi_{\alpha ,n} (x,t;E) = \phi_{\alpha , n} (x;E) e^{-i Et/\hbar}
	\sum_{l =-\infty}^\infty J_l 
	\left( \frac{eV_\alpha}{\hbar\omega}\right) 
	e^{-il\omega t}
\label{res}	
\end{equation}
where $\phi_{\alpha, n} (x;E) $ is the wave function describing an 
incoming (or outgoing) carrier in contact $\alpha$ in channel $n$
in the absence of a modulation potential,   
and $J_l$ is the $l$'th order Bessel function. 
Thus the potential modulation leads for each state with 
central energy $E$ to side bands at energy $E+ l\hbar \omega$
describing carriers which have absorbed $l > 0$ modulation 
quanta or have emitted $l < 0$ modulation quanta $\hbar \omega$.
Here we have assumed that all potentials oscillate in phase.
If one allows for a different phase $\phi_\alpha$ for each contact
$\alpha$ that will add a term $e^{-il\phi_\alpha}$ to each term in the sum in the 
wave-function above. Below, for simplicity, we assume that all contact potentials
are in phase.

We now suppose that the modulation potential exists only far away
from the conductor and that the modulation potential vanishes as 
we approach the conductor.
Thus there is a transition region 
from a portion of the lead in which the 
potential is oscillating and a portion of the lead 
close to the conductor where we initially assume 
that the potential is time-independent and 
equal to the equilibrium potential. 
Now we need the wave function in the time-independent potential 
region. This leads to a matching problem. 
If the transition is adiabatic a state with energy $E$ in the conductor
obtains a contribution from all reservoir states  
with central energy 
$E-l\hbar\omega$ due to its side band of amplitude 
$J_l\left(\frac{eV_\alpha}{\hbar\omega}\right)$ at energy $E$. 
In the notation of second quantization 
the annihilation operator of an incoming state close to the conductor
is
\begin{equation} 
    \hat{\bf a}_{\alpha,n} (E) = \sum_l 
    \hat{\bf a}'_{\alpha , n} (E-l\hbar\omega)
	J_l \left(\frac{eV_\alpha}{\hbar\omega}\right) . \label{lead}	
\end{equation}
up to corrections of the order of $\hbar \omega /E_F$ which arise 
from the difference of the wave vectors of the 
sidebands $p_{l} = \sqrt{2m (E + l\hbar \omega)}/\hbar$ and 
the wavevector at energy $E$.
The current operator Eq.~(\ref{current_op})
is expressed in terms of the incoming (and outgoing) states 
of the stationary time-independent scattering problem. 
Eq.~(\ref{lead}) can now be used to 
find the current operator in terms of the reservoir states 
$\hat{\bf a}'_{\alpha , n}$. 
The current operator becomes
\begin{eqnarray}
    \hat{I}_\alpha(t) &=& \frac{e}{h} \int dE \int dE' \sum_{\gamma\delta }
	\sum_{lk=-\infty}^\infty
	(\hat{\bf a}')_\gamma^\dagger(E-l\hbar\omega) \nonumber \\
	&& J_l\left(\frac{eV_\gamma}{\hbar\omega}
	\right) J_k\left(\frac{eV_\delta}{\hbar\omega}\right) 
	e^{i(E-E')t/\hbar} \nonumber \\
	&& {\bf A}_{\gamma\delta}(\alpha,E,E')
	\hat{\bf a}'_\delta(E'-k\hbar\omega) \label{currentop}
\end{eqnarray}
where we have introduced the {\em current matrix} \cite{buttiker92}
\begin{equation}
    {\bf A}_{\delta\gamma}(\alpha,E,E^{\prime}) = \delta_{\alpha\delta}
    \delta_{\alpha\gamma} {\bf 1}_\alpha - 
    {\bf s}_{\alpha\delta}^\dagger(E) {\bf s}_{\alpha\gamma}(E^{\prime}).
\end{equation}

It is assumed that the modulation imposed on the system is so slow that
the contacts can still be regarded as being in a dynamic equilibrium state.
Thus the quantum statistical average can be 
found by evaluating averages of the $\hat{\bf a}_\alpha(E-l\hbar\omega)$
as for an equilibrium system. 
Replacing the $\hat{\bf a}_\alpha(E-l\hbar\omega)$
by their equilibrium statistical expectation values we find, 
\begin{eqnarray} 
    I_\alpha(t) &=& \frac{e}{h} \int dE \sum_{\gamma,lk} \label{newcurrent} 
	\mbox{Tr} {\bf A}_{\gamma\gamma}(\alpha,E,E+(k-l)\hbar\omega) 
	\times \\
	&& J_l\left(\frac{eV_\gamma}{\hbar\omega}\right)
	J_k\left(\frac{eV_\gamma}{\hbar\omega}\right)
	e^{-i(k-l)\omega t} f_\gamma(E-l\hbar\omega) . \nonumber	 
\end{eqnarray}
where $f_\gamma(E)=f(E-\mu_\gamma)$ 
is the Fermi distribution function for contact $\gamma$.
Here $\mu_\gamma$ is the electrochemical potential of reservoir
$\gamma$. 
In Eq.~(\ref{newcurrent}) the trace is over all channels in lead $\alpha$.
Taking into account the symmetry properties of the current matrix 
under exchange of the energy arguments it can be shown that the 
current given by Eq.~(\ref{newcurrent}) is real. 

From Eq.~(\ref{newcurrent}) we find that for the dc-current only 
the terms $l = k$ contribute. In this case, as is seen by looking 
at Eq.~(\ref{newcurrent}), the energy arguments of the 
current matrix are equal. The trace of the current matrix
at equal energy arguments and equal lower lead indices 
are just transmission and reflection 
probabilities. We define $T_{\alpha\gamma}(E) = -
\mbox{Tr} {\bf A}_{\gamma\gamma}(\alpha,E,E)$. 
For unequal indices $\alpha$ and $\gamma$ this is 
the transmission probability 
for carriers incident in lead $\gamma$ to be transmitted 
into contact $\alpha$.  
If also $\alpha  = \gamma$ the trace of the current matrix 
is equal to  the probability $R_{\alpha\alpha}$ of carriers incident in 
lead $\alpha$ to be reflected 
back into lead $\alpha$, minus  
the number of quantum channels $N_{\alpha}$ at energy $E$.  
In this notation, particle conservation 
in the scattering process is expressed by the sum rule 
$\sum_{\gamma} T_{\alpha\gamma} = 0$. For the dc-current we find thus 
\begin{eqnarray} 
    I_\alpha(0) = - \frac{e}{h} \int dE \sum_{\gamma,l}
	T_{\alpha\gamma}(E) 
	J^{2}_l\left(\frac{eV_\gamma}{\hbar\omega}\right)
	f_\gamma(E-l\hbar\omega) . 
\label{cur0}	 
\end{eqnarray}
Now we expand in this expression 
the Bessel functions in powers of the applied oscillating potentials
$V_\gamma$. The zero-th order terms gives the dc-current 
$I^{dc,(0)}[\{V_{\beta}\}]$ that flows as a consequence of stationary 
differences in the applied potentials. We 
use a superscript $(0)$ to denote a response to an external potential only.
Since the potential in the interior is kept
fixed this $I-V$ characteristic is not gauge 
invariant. A discussion is 
provided in Ref.~\cite{buttiker93} and 
by Christen and one of the authors \cite{christen96}. The next term is second order 
in the amplitudes of the oscillating voltages. For identifying conductance
coefficients recall that the applied potential is of the form
$V_\gamma(t)=\frac{1}{2} V_\gamma(\omega) e^{i\omega t}+
\frac{1}{2} V^*_\gamma(\omega) e^{-i\omega t}$,
with the amplitudes taken as real. Thus, from the calculated response to $V_\gamma(t)$ we
need to extract the response to the Fourier amplitudes.
These second order terms
are determined by the photo-conductances 
\begin{eqnarray}
    g_{\alpha\beta\gamma}^{(0)}[\omega,-\omega;\{V_\delta(0)\}] &=& 
         - \delta_{\beta\gamma}  \frac{e^3}{h}
	\int dE {T}_{\alpha\beta}[E;\{V_\delta(0)\}] \label{photoconductance} \\
	&& \hspace*{-0.6cm} \frac{f_\beta(E+\hbar\omega)+f_\beta(E-\hbar\omega)-2f_\beta (E)}
	{(\hbar\omega)^2}. \nonumber
\end{eqnarray}
The photo-conductance 
$g_{\alpha\beta\gamma}^{(0)}[\omega,-\omega;\{V_\delta(0)\}]$ 
determines the zero-frequency current in contact 
$\alpha$ in response to a second order voltage oscillation 
$V^{2}_\beta(\omega)$ at contact $\beta$. Note that the external 
photo-conductance generated by bilinear products $V_\beta(\omega) 
V_\gamma(\omega)$ with $\beta$ unequal to $\gamma$ vanishes. 
Instead of a second order difference in Fermi functions 
we can express the photo-conductance as a second order difference of 
transmission probabilities
\begin{eqnarray}
    g_{\alpha\beta\gamma}^{(0)}[\omega,-\omega] &=& 
         - \delta_{\beta\gamma}  \frac{e^3}{h} 
	\int dE f_\beta (E) \times 	\label{photoconductance1} \\
	&& \hspace*{-0.25cm} \frac{T_{\alpha\beta}(E+\hbar\omega)+
	        T_{\alpha\beta}(E-\hbar\omega)-
	        2 T_{\alpha\beta}(E)}{(\hbar\omega)^2} .\nonumber
\end{eqnarray}
For simplicity we have not explicitly indicated the dependence
on the stationary potentials ${V_\delta(0)}$.
Eq.~(\ref{photoconductance1}) shows clearly that we obtain an externally induced photo-current
only if the transmission probabilities through the sample
are energy dependent. 
Thus, for a quantum point contact or for a quantized Hall conductor,
where we encounter situations characterized by transmission
probabilities which are either zero or one, there is 
no externally induced photo current. 
%We should also check the literature
%whether their are papers which agree or disagree with this 
%important statement. 
%Look at papers by Anna Grincwaig, Shekter and Jonson\cite{grincwaig}. 
This form of the photo-conductance also makes it evident that 
current conservation is satisfied due to the unitarity of the scattering 
matrix: The sum of all photo-conductances over all contacts adds up to 
zero, 	$\sum_{\alpha} g^{(0)}_{\alpha\beta\gamma}[\omega,-\omega] = 0$. 
However, similar to the dc I-V characteristic these conductances are not 
gauge invariant. 
The sum $\sum_{\beta } g^{(0)}_{\alpha\beta\gamma}[\omega,-\omega] $
does not vanish and consequently the photo-current evaluated with 
these expressions depends not only on voltage differences. 

Let us next consider the displacement current. 
The current at the frequency $\omega$ is determined by 
the terms in Eq.~(\ref{newcurrent}) for which $k - l =1$ and 
it is given by 
\begin{eqnarray} 
    I_\alpha(\omega) &=& \frac{e}{h} \int dE \sum_{\gamma,l}
	\mbox{Tr} {\bf A}_{\gamma\gamma}(\alpha,E,E+\hbar\omega) 
	\times \nonumber \\
	&& J_l\left(\frac{eV_\gamma}{\hbar\omega}\right)
	J_{l+1}\left(\frac{eV_\gamma}{\hbar\omega}\right)
	f_\gamma(E-l\hbar\omega) . \label{discurrent} \label{cur1}	 
\end{eqnarray}
Linearising the response to an oscillating external potential yields
the admittance previously found \cite{buttiker93,pretre},  
\begin{eqnarray}
    g_{\alpha\beta}^{(0)}[\omega;\{V_\gamma(0)\}] &=& \frac{e^2}{h} \int dE 
    \mbox{Tr} {\bf A}_{\beta\beta}[\alpha,E,E+\hbar\omega;\{V_\gamma(0)\}] 
    \nonumber \\
	&& \times \frac{f_\beta (E)-f_\beta (E+\hbar\omega)}{\hbar\omega}.	
\label{accond}	
\end{eqnarray}
The external admittance given by Eq.~(\ref{accond}) has been the 
starting point of a self-consistent discussion of ac transport 
based on the scattering matrix approach. The approach has been illustrated 
in a number of works \cite{christenprl,brouwerepl,guo}. 
The next term in the expansion is third order in the 
oscillating potentials and will not be needed here. 
  
We remark that the external photo-conductances Eq.~(\ref{photoconductance1})
are like the dc-current determined by transmission probabilities
only. In contrast, the displacement current invokes products
of scattering matrices at different energies and thus depends 
also on the phases of the scattering matrix. Expressed in a more 
physical language, the displacement current is sensitive to the 
densities of carriers, expressed here via energy derivatives of 
phases. Below we find that 
the self-consistent photo-current contains in fact not only
transmission probabilities but, like the displacement current, also
information on the charge accumulated in the conductor. 

Before considering the effect of screening, 
we discuss the relation of the external response to 
previous work. A discussion of shot noise in a conductor with applied
ac voltages can be found in Appendix A.

\subsection{Two-terminal conductors}

We consider a two-terminal conductor which consists of 
a tunneling barrier connected on either side to a large 
contact. This is the geometry considered by Tien and Gordon \cite{tien}
and Tucker \cite{tucker}.  
The results we obtain from the external response 
described above are in agreement with these earlier works. 
Below we emphasize the arbitrary nature of these results 
arising from the lack of a selfconsistent treatment of the Coulomb interaction. 
As a consequence, different
but physically identical configurations of voltages lead to different
results. Later we will show how the results 
for these two arrangements change when screening
is taken into account, for the specific example where the barrier 
is a resonant tunneling barrier.

First we consider the Tien and Gordon case, 
where one of the contact potentials is oscillating 
and the
other is kept fixed, $V_1(\omega)=V(\omega)$ and $V_2(\omega)=0$. 
For simplicity we assume that the scattering matrix has been 
diagonalized such that transmission through 
the barrier is described by a transmission probability $T_m(E)$
and a reflection probability $R_m(E)$ for the m-th eigen channel. 
Using Eq.~(\ref{cur0}) and using the sum rule for Bessel functions,
$\sum_l J_{l+k}(x) J_l(x)=\delta_{k0}$,
we find 
\begin{eqnarray}
    I_1(0) &=& - \frac{e}{h} \int dE \sum_{lm} 
	J_l^2\left(\frac{eV(\omega)}{\hbar\omega}\right)
	T_m(E) \times \nonumber \\
	&& \times [ f_1(E+l\hbar\omega) - f_2(E)]. \label{tiengordon}
\end{eqnarray}
Tien and Gordon \cite{tien} express the transmission 
probabilities with the help of Bardeen's formula 
$T = 4\pi|t|^{2}\nu_{1}(E) \nu_{2}(E+l\hbar\omega)$
in terms of a matrix element $t$ and 
the density of states $\nu_{1}(E)$ and $\nu_{2}(E)$
to the left and right of the barrier. 
In our work the energy $E$ is a global variable, whereas 
Tien and Gordon measure energy in the densities of states 
from the conduction band bottom to the left and right 
of the barrier.

%\subsubsection{Barrier oscillates}

%The second example again considers a tunneling
%barrier but the contact potentials are kept fixed and the barrier
%potential oscillates. 

The time-dependent current was investigated by Tucker for the same
geometry \cite{tucker}.
Using Eq.~(\ref{newcurrent}) we find
\begin{eqnarray}
    I_1(t) &=& \sum_{lk\gamma} \mbox{Tr} {\bf A}_{\gamma\gamma}(\alpha,E,E+k\hbar\omega)
	\nonumber \\ && 
	J_l\left(\frac{eV(\omega)}{\hbar\omega}\right)
	J_{l+k}\left(\frac{eV(\omega)}{\hbar\omega}\right) 
	e^{-ik\omega t}\times \nonumber \\
	&& \times f_\gamma(E-l\hbar\omega).
\end{eqnarray}
This is the result of Tucker \cite{tucker} except that he considers 
the barrier transmission to be energy independent. 

However, the results obtained above are not invariant under an equal 
shift of all potentials. 
For example, in the Tien and Gordon case, an experimentally equivalent situation
would be to set $V_1(\omega)=V(\omega)/2$ and 
$V_2(\omega)=-V(\omega)/2$. This, however, yields a
different result in the non-interacting theory. Even worse,
setting $V_1(\omega)=V_2(\omega)=V(\omega)/2$ should 
yield no photo-current, but gives the same as for 
$V_1(\omega)=V(\omega)/2$ and $V_2(\omega)=-V(\omega)/2$.
To remedy this we 
introduce in the next section
a simple selfconsistent scheme to achieve charge and current 
conservation, similar to one used previously \cite{buttiker93,pedersen}.

\subsection{Density operator}

When applying voltages to the conductor, the sample will be charged.
The net charge of the sample in response to a potential applied to a
contact can be decomposed into two contributions:
A charge response, called the injectance of 
the contact, at fixed internal electric potential
and a charge response due to an electrically induced potential. 
Here we determined the injectances
of a multi-terminal conductor. 
In the next section these results are used when treating the problem with
interactions.

At zero frequency, the number of electrons in the sample is determined by the operator
\cite{buttikermathphys}
\begin{eqnarray}
    \hat{N} &=& \sum_{\alpha\beta nm} \int d^2{\bf r} \int dE \,
	\nu^{1/2}_{\alpha n}(E) \nu^{1/2}_{\beta m}(E) \times \\
	&& \Psi^*_{\alpha n}(r,E) \Psi_{\beta m}(r,E)
	\hat{a}_{\alpha n}^\dagger(E) \hat{a}_{\beta m}(E), \nonumber
\end{eqnarray}
where $\nu_{\alpha n}(E)$ is the density-of-states for channel $n$ in
contact $\alpha$, and $\Psi_{\alpha n}(r,E)$ is the corresponding
wave-function for a scattering state describing carriers incident 
in contact $\alpha$ in channel $n$.

We now define the partial density-of-states matrix 
$\frac{d {{\bf \cal N}}_{\alpha\beta}}{dE}$, 
with elements
\begin{equation}
    \frac{d {\cal N}_{\alpha\beta,nm}}{dE} = 
    \int d^2{\bf r}\, \nu^{1/2}_{\alpha n}(E) \nu^{1/2}_{\beta m}(E) 
	\Psi^*_{\alpha n}(r,E) \Psi_{\beta m}(r,E).
\end{equation}
This matrix can also be expressed in terms the scattering matrix and its
derivatives \cite{buttikermathphys}
\begin{equation}
    \frac{d {\bf {\cal N}}_{\gamma\delta}}{dE} = 
    -\frac{1}{4\pi i} \sum_\beta \left[
    {\bf s}^\dagger_{\beta\gamma}(E) \frac{d {\bf s}_{\beta\delta}(E)}{dE}
    - \frac{d {\bf s}^\dagger_{\beta\gamma}(E)}{dE} {\bf s}_{\beta\delta}(E) \right].
\end{equation}
Using this and Eq.~(\ref{lead}) we find the number operator in the
presence of oscillating contact potentials
\begin{eqnarray}
    \hat{N} &=& \sum_{\alpha\beta lk} \int dE \, J_l\left(
	\frac{eV_\alpha}{\hbar\omega}\right) J_k\left(
	\frac{eV_\beta}{\hbar\omega}\right) \times \nonumber \\
	&& (\hat{\bf a}')_\alpha^\dagger(E-l\hbar\omega)
	\frac{d {\bf {\cal N}}_{\alpha\beta}}{dE}
	\hat{\bf a}'_\beta(E-k\hbar\omega),
\end{eqnarray}
with the expectation value
\begin{equation}
    N = \sum_{\alpha l} \int dE \, J_l^2\left(
	\frac{eV_\alpha}{\hbar\omega}\right) \mbox{Tr}
	\frac{d {\bf {\cal N}}_{\alpha\alpha}}{dE} 
	f_\alpha(E-l\hbar\omega).
\end{equation}
We can shift the frequency dependence from the Fermi function 
to the partial density of states 
\begin{equation}
    N = \sum_{\alpha} \int dE \, J_l^2
        \left(\frac{eV_\alpha}{\hbar\omega}\right) 
        \frac{d N^{(0)}_\alpha}{dE}
	f_\alpha(E).
\end{equation}
and thus identify the injectance $\frac{d N^{(0)}_{\alpha}}{dE}$
at energy $E$ in the presence of a potential variation at contact $\alpha$.
Here the upper index $0$ is once more used to emphasize that this density is evaluated 
at fixed internal potential.   
To second
order in the oscillating potential, $V_{\alpha}(\omega)$,
the injectance is 
%\begin{eqnarray}
%    \frac{d N^{(0)}_\alpha}{dE} &=& \mbox{Tr} \left[
%	\left(1-\left(\frac{e}{\hbar\omega}\right)^2 |V_\alpha(\omega)|^2 \right)
%	\frac{d {\bf {\cal N}}_{\alpha\alpha}(E)}{dE} \right. \label{injectance} \\
%	&& 
%	+ \frac{1}{2}\left(\frac{e}{\hbar\omega}\right)^2 |V_\alpha(\omega)|^2 
%	\times \nonumber \\
%	&& \left. \left( \frac{d {\bf {\cal N}}_{\alpha\alpha}(E+\hbar\omega)}{dE}+
%	\frac{d {\bf {\cal N}}_{\alpha\alpha}(E-\hbar\omega)}{dE} \right) 
%	\right] \nonumber.
%\end{eqnarray}
\begin{eqnarray}
    \frac{d N^{(0)}_\alpha}{dE} &=& \mbox{Tr} \left[
	\frac{d {\bf {\cal N}}_{\alpha\alpha}(E)}{dE} 
	 - \frac{e^2}{2} 
	\frac{ |V_\alpha(\omega)|^2 }{(\hbar\omega)^2}
	\times \right.  \label{injectance} \\
	&& \hspace*{-0.7cm} \left. \left(
	\frac{d {\bf {\cal N}}_{\alpha\alpha}(E+\hbar\omega)}{dE}+
	\frac{d {\bf {\cal N}}_{\alpha\alpha}(E-\hbar\omega)}{dE} -
	2\frac{d {\bf {\cal N}}_{\alpha\alpha}(E)}{dE} \right)
	\right] \nonumber 
\end{eqnarray}
In the limit that $|V_{\alpha}(\omega)|$ becomes small compared to $\hbar \omega$ 
the injectance is that produced by a static 
voltage.

\section{Internal Response: Self-consistent screening}

In response to a potential variation at a contact the charge distribution 
in the interior of the sample is driven away from its equilibrium pattern. 
Coulomb interactions oppose such a variation. 
In the problem of interest here a variation of the 
sample charge can come about both because we in general 
consider a biased sample such that a dc-current flows and because 
we subject the sample to ac-voltages. 
In general it is a non-equilibrium dynamical potential landscape that 
matters. Here for simplicity we consider the sample 
to be zero-dimensional and assume that it suffices to consider 
a single internal potential $U$. 
Such an approximation is often used 
in the literature on the Coulomb blockade and in the scattering 
approach to electrical conduction has been 
used to discuss the non-linear I-V characteristic
of mesoscopic samples \cite{christen96}
and ac-transport in Refs.~\cite{pedersen,pretre,buttiker94}.
At equilibrium, if all voltages
at the contact of the sample are equal, and in the absence 
of ac-potentials, the value of this potential is $U = U_{eq}$.
Our first task is to determine the zero-frequency part of
this potential.

To be more specific we now consider a sample coupled to a gate,
as an example see Fig.~1.
We denote the contact to the gate 
by the index $g$ and the capacitance of the central 
region of the conductor to the gate by $C_g$. 
The capacitance between the central region of the conductor 
to the reservoir $\alpha$ is denoted by $C_{\alpha}$. 
Next we introduce an index $\nu$ which runs over all
$N$ current contacts of the sample $ \nu = \alpha = 1, 2,   ..N$
and in addition includes the contact to the gate $\nu = N+1=g$.

\subsection{Static internal potential}

Consider first the equilibrium potential $U_0^{eq}$. 
The grandcanonical potential with the Coulomb
energy included is minimal for a potential $U_0^{eq}$
that obeys the Poisson equation. 
In our case the Poisson equation is discretisized 
and is expressed with the help of the geometrical capacitances 
introduced above.
The net electronic charge on the sample is that permitted by 
the Coulomb interaction: 
\begin{equation}
    Q- Q^{+} = \sum _{\nu} C_{\nu} (U - V_\nu) 
    \label{pq}
\end{equation}
Here $Q$ is the electronic charge, $Q^{+}$ is an effective "ionic 
charge" created by the donors and $C_{\nu}$ 
are the geometrical capacitances. 

For $V_{\nu} = 0$, the equilibrium charge $Q = Q^{eq}_0$ and the equilibrium 
potential $U = U^{eq}_0$ follow from Eq. (\ref{pq}) as follows. 
The electronic charge on the conductor can be expressed as a sum of all the 
charges injected from the various contacts, 
\begin{equation}
    Q^{eq}_0 = \sum_\alpha \int_{-\infty}^{\mu} e \frac{d N_\alpha(U_0^{eq})}{dE} dE ,
    \label{ueq} 
\end{equation} 
where the injectance \cite{buttiker94,buttiker96} of contact $\alpha$ is given by  
Eq.~(\ref{injectance}).
Note that the scattering matrix and thus the injectance 
also depends on $U_0^{eq}$. Eq.~(\ref{ueq}) is thus a self-consistent 
equation for the equilibrium potential. 

Next, let us keep the ac-voltages turned of but 
apply dc voltages to the contacts, charge will flow into the
conductor causing a shift of the static potential in the barrier. 
We denote the resulting potential by $U_0$. 
It is a function of the applied potentials $V_\nu$ since now 
the injected charge depends on all the applied voltages.
The injected charge is given by
\begin{equation}
    Q_0 = \sum_\alpha \int_{-\infty}^{\mu_{\alpha}} e \frac{d N_\alpha(U_{0})}{dE} dE ,
    \label{qneq} 
\end{equation} 
Using Eq.~(\ref{ueq}) to express the effective background charge 
in terms of the scattering matrix and the charges on the capacitors gives
the following self-consistent
equation for determining the internal static potential in the sample
\begin{eqnarray} 
    && \sum_\alpha \int_{-\infty}^{\mu_\alpha} e \frac{d N_\alpha(U_0)}{dE} dE -
	\int_{-\infty}^{\mu_{eq}} \sum_\alpha e \frac{d N_\alpha(U_0^{eq})}{dE} dE \nonumber \\
	&& = \sum_\nu C_\nu (U_0 - V_\nu) \label{charge} .	
\end{eqnarray}
Here $V_{\alpha} =\mu_{\alpha} -\mu$
is the deviation of the electrochemical potential in contact 
$\alpha$ from its equilibrium value $\mu$. 
This approach was used by Christen and B\"uttiker \cite{christen96} to study
the nonlinear conductance for a resonant tunneling barrier.

Next, consider next the case that 
is really of interest here. In addition to possible static voltage differences 
we have time-dependent potentials at the 
contacts. As a consequence the (unscreened) charge $Q(t)$  
in the sample is also a function of time. It can be Fourier transformed, 
and we expect Fourier components at the oscillation frequency $\omega$ 
of the voltage and at all higher harmonics, $k\omega$. 
As a consequence the potential inside the conductor will also 
oscillate and will similarly have Fourier components 
at all harmonics, $U(k\omega)$. 
If an oscillating voltage at a contact, 
due to non-linear processes, also changes the time-averaged charge in 
the sample then the potential $U_0$ as determined above would be 
modified by the presence of the oscillating potentials. To take this into
account we write the injected charge as a response to external potentials
in the presence of a self-consistently determined static potential
plus the response from the internal oscillating potential. The response to the
internal potential is determined by three unknown response coefficients,
$\chi_{i\alpha}$, $\chi_{\alpha i}$ and $\chi_{ii}$ such that,
\begin{eqnarray}
    Q_0 &=& \sum_\alpha \int 
    e\frac{d N^{(0)}_\alpha}{dE}dE + 
	\sum_\alpha \int dE \chi_{i\alpha}(E) U^*(\omega) V_\alpha(\omega) \nonumber \\
	&& + \sum_\alpha \int dE \chi_{\alpha i}(E) V^*_\alpha(\omega) U(\omega) +
	\chi_{ii}(E) | U(\omega) |^2.
\end{eqnarray}
To determine $\chi_{i\alpha}$, $\chi_{\alpha i}$ and $\chi_{ii}$ we use the fact 
that the injectance be invariant under a shift of all oscillating potentials
by an equal amount. This yields the coefficients
\begin{eqnarray}
    \chi_{i\alpha}(E) &=& \chi_{\alpha i}(E) = \mbox{Tr} \left[
	\frac{1}{2}\left(\frac{e}{\hbar\omega}\right)^2
	\left( \frac{d {\bf {\cal N}}_{\alpha\alpha}(E+\hbar\omega)}{dE}
	\right. \right. \nonumber \\
	&& + \left. \left.
	\frac{d {\bf {\cal N}}_{\alpha\alpha}(E-\hbar\omega)}{dE} 
	-2\frac{d {\bf {\cal N}}_{\alpha\alpha}(E)}{dE} 
	\right) \right] ,\\
    \chi_{ii}(E) &=& -\sum_\alpha \chi_{i\alpha}(E).
\end{eqnarray}
With this we can express the gauge-invariant injectance as
\begin{eqnarray}
    \frac{d N_\alpha}{dE} &=& \mbox{Tr} \left[
	\frac{d {\bf {\cal N}}_{\alpha\alpha}(E)}{dE} 
	 - \frac{e^2}{2} 
	\frac{|V_\alpha(\omega)-U(\omega)|^2 }{(\hbar\omega)^2}
	\times \right. \label{inj} \\
	&& \hspace*{-0.7cm} \left. \left(
	\frac{d {\bf {\cal N}}_{\alpha\alpha}(E+\hbar\omega)}{dE}+
	\frac{d {\bf {\cal N}}_{\alpha\alpha}(E-\hbar\omega)}{dE} -
	2\frac{d {\bf {\cal N}}_{\alpha\alpha}(E)}{dE} \right) \nonumber
	\right] 
\end{eqnarray}

Eqs.~(\ref{charge}) and (\ref{inj}) now allows us to find the static internal
potential $U_0$ in the presence of static and oscillating contact voltages.
Note that $\frac{d N_\alpha}{dE}$ depends on $U_0$ 
since the scattering matrix depends on $U_0$. The potential $U_0$
depends on the dc-voltages applied to the sample and depends through
non-linear processes on the amplitudes of the ac-voltages and the frequency. 

Our next task is now to find the {\em current} response to the 
oscillating internal potential $U(t)$.

\subsection{dc-current}

Consider first the photo-induced dc-current. 
The dc-current can be divided into two parts, one due to direct transmission
processes, and one due to transmission after absorption (emission) of a photon
followed by its emission (absorption). 
Both processes take place in a self-consistently determined 
electrostatic background, which depends on all voltages at all frequencies.
\begin{equation}
    I_\alpha(0) = I_{\alpha}^{dc}[\{V_\beta(0)\}]
	+ I^{ph}_\alpha[\{V_\beta(0)\}].
\end{equation}
Here $I_{\alpha}^{dc}[\{V_\beta(0)\}]$ is determined from the first term of
the sum in Eq.~(\ref{cur0}), where now the scattering matrix depends on $U_0$.

The photo-current can be
written generally as the sum of the response to the external oscillating potential and
the internal potential $U(\omega)$. 
To proceed we now consider $\epsilon = eU(\omega)/(\hbar\omega)$,
a small parameter in which we can expand. All the oscillating
contact potentials are also of order $\epsilon$.  
In this work we will stop this expansion at the first 
non-trivial order. Since photon-assisted tunneling is of second
or higher order in the oscillating potentials, we carry the expansion to second order.

For the photo-current we obtain 
\begin{eqnarray}
     I^{ph}_\alpha[\{V_\beta(0)\}] &=& \sum_{\beta\gamma} 
	g_{\alpha\beta\gamma}^{(0)}[\omega,-\omega;\{V_\delta(0)\}] V_\beta(\omega)
	V_\gamma^*(\omega) \nonumber \\
	&& + \sum_\beta g_{\alpha\beta i}[\omega,-\omega;\{V_\delta(0)\}] V_\beta(\omega)
	U^*(\omega) \nonumber \\
	&& + \sum_\gamma g_{\alpha i\gamma}[\omega,-\omega;\{V_\delta(0)\}]
	U(\omega) V_\gamma^*(\omega) \nonumber \\
	&& + g_{\alpha ii}[\omega,-\omega;\{V_\delta(0)\}] U(\omega) U^*(\omega)
	\label{full}
\end{eqnarray}
where the index $i$ refers to responses due to the internal potential 
$U(\omega)$.

The responses to the internal potential are found by demanding that the
current is invariant with respect to a shift of all voltages
(gauge invariance). Lowering
all voltages at frequency $\omega$ by $U(\omega)$ 
shifts the internal potential to the external 
potentials. Comparing the resulting expression with Eq.~(\ref{full})
gives 
\begin{eqnarray}
    g_{\alpha\beta i}[\omega,-\omega;\{V_\delta(0)\}] &=& -\sum_\gamma 
	g_{\alpha\beta\gamma}^{(0)}[\omega,-\omega;\{V_\delta(0)\}], \\
    g_{\alpha i\gamma}[\omega,-\omega;\{V_\delta(0)\}] &=& -\sum_\beta 
	g_{\alpha\beta\gamma}^{(0)}[\omega,-\omega;\{V_\delta(0)\}], \\
    g_{\alpha ii}[\omega,-\omega;\{V_\delta(0)\}] &=& \sum_{\beta\gamma} 
	g_{\alpha\beta\gamma}^{(0)}[\omega,-\omega;\{V_\delta(0)\}].
\end{eqnarray}
Thus the photo-responses to the internal potential are determined 
by combinations of external photo-conductances. 

With these conductances, the dc-current can be written as 
\begin{eqnarray}
    I_\alpha(0) &=& I_\alpha^{dc}[\{V_\beta(0)\}] \label{current} \\
	&& +
	\sum_\beta g_{\alpha\beta\beta}^{(0)}[\omega,-\omega; \{V_\gamma(0)\}] 
	|V_\beta(\omega)-U(\omega)|^2. \nonumber
\end{eqnarray}
Note that the current depends on the difference between an applied voltage and
the internal voltage only.

All the non-linear transport coefficients in Eqs.~(\ref{full}-\ref{current}) also
depend on $U_0$, the self-consistent dc-potential.

\subsection{Displacement current}

The current at frequency $\omega$ is only needed to first order in the
applied oscillating voltages. In addition to the external 
potential the oscillating internal potential 
also contributes to the current. In the presence of the internal potential 
the general form for the current is to first order in the potentials 
\begin{eqnarray}
    I_\alpha(\omega) &=& 
	\sum_\beta g_{\alpha\beta}^{(0)}[\omega;\{V_\gamma(0)\}] 
	V_\beta(\omega) \nonumber \\
	&& + g_{\alpha i}[\omega;\{V_\gamma(0)\}] U(\omega) .
\end{eqnarray}
Here $g_{\alpha\beta}^{(0)}[\omega;\{V_\gamma(0)\}]$ are the 
external ac-conductances given by Eq.~(\ref{accond})
and $g_{\alpha i}$ is the ac-response to the internal potential.  
Again we determine $g_{\alpha i}$
through the requirement that this expression is invariant 
under an overall shift of the potential. 
This gauge invariance argument determines the response 
to the internal potential 
in terms of external responses;
$g_{\alpha i}[\omega;\{V_\gamma(0)\}]=
-\sum_\beta g_{\alpha\beta}^{(0)}[\omega;\{V_\gamma(0)\}]$.

Both for the dc-current and the ac-current we now know 
the response to the external voltages $V_\gamma(0)$ and 
to the internal potential $U(t)$. But the internal 
potential is thus far not determined. This is our next task. 

To be more specific we now return to the sample shown in Fig.~\ref{structure}.
The current at contact $\alpha$ is the particle current plus the
displacement current (capacitive) current $I_{\alpha} (\omega) - i\omega 
C_{\alpha} (V_{\alpha}(\omega) - U(\omega))$ with $I_{\alpha}$ as determined above. 
The current from the gate to the sample is purely capacitive and 
is given by $I_g(\omega) = -i\omega (V_{g}(\omega) - U(\omega))$. 
Since the overall charge at frequency $\omega$ 
is conserved the sum of these currents must 
vanish. Thus we must have 
\begin{equation}
\sum_\alpha I_\alpha(\omega) = -i\omega \sum_\nu C_\nu (U(\omega)-V_\nu(\omega)).
\end{equation}
Solving this equation for the internal potential yields 
\begin{equation}
    U(\omega) = \frac{\sum_{\alpha\beta} g_{\alpha\beta}^{(0)}
    [\omega;\{V_\gamma(0)\}]
	V_\beta(\omega)-i\omega \sum_\nu C_\nu V_\nu(\omega) }
	{\sum_{\alpha\beta} 
	g_{\alpha\beta}^{(0)}[\omega;\{V_\gamma(0)\}]
	 - i\omega \sum_\nu C_\nu}.
	 \label{uself}
\end{equation}
The external ac-conductances and the geometrical capacitances determine 
the potential $U(\omega)$ and determine the self-consistent dc-current 
due to photo-assisted tunneling and the self-consistent ac-conductances.

\section{Resonant tunneling barrier}

As an application of the selfconsistent theory developed above, we consider
the photo-induced dc-current through a resonant tunneling barrier. The 
experimental setup is taken as sketched in Fig.~\ref{structure}. Each side
of the resonant barrier is connected to reservoirs with chemical potentials
$\mu_1$ and $\mu_2$ and capacitances $C_1$ and $C_2$.
The interior of the barrier is coupled to a gate with a capacitance
$C_g$. For simplicity we assume that the gates are macroscopic with no
dynamics of their own. A dc bias will be applied by making 
$eV \equiv e(V_{1} - V_{2}) = \mu_1 - \mu_2 $
non-zero.

The scattering matrix close to a resonance is given by the Breit-Wigner
formula \cite{buttiker88,breit,landau}
\begin{equation}
\label{bw}
    s_{mn} = \left[ \delta_{mn} - i \frac{\sqrt{\gamma_m\gamma_n}}
	{E-E_0-eU_0+i\gamma/2}
	\right] e^{i(\delta_m+\delta_n)} .
\end{equation}
Here $\gamma_n$, $n = 1, 2$ are the partial widths of the resonance 
proportional to the tunneling probability through the left 
and right
barrier and $\gamma=\sum_n \gamma_n$ 
is the total width of the resonance. 
$\delta_m$ are the phases acquired in the
reflection or transmission process and $E_0$ is the position of the resonance.
The term $eU_0=e\frac{V_1(0)+V_2(0)}{2}+W$ ensures
invariance upon a shift of the dc voltages\cite{christen96}. $W$ is determined
by the condition Eq.~(\ref{charge}) and Eq.~(\ref{inj}),
and is a function of $V_1(0)-V_2(0)$ only.
The injectivities are \cite{buttiker94}
\begin{equation}
   \frac{dN_\alpha}{dE}=\frac{1}{2\pi}\frac{\gamma_\alpha}{(E-E_0-eU_0)^2+(\gamma/2)^2}.
\end{equation}

The Breit-Wigner formula is a reasonable form for the scattering matrix as long as
the energy does not get close to the next resonance level. Assuming that the level
spacing of our system is large enough such that neighboring levels can safely be ignored
we will use the formula in a wide energy range.

\begin{figure}
\narrowtext
\epsfysize=9cm
\epsfxsize=7cm
\centerline{\epsffile{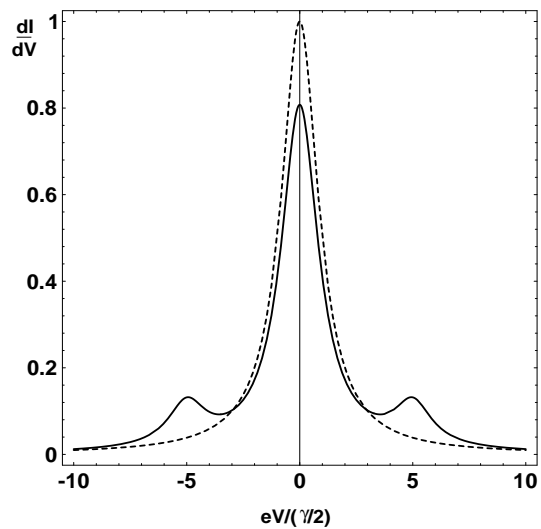}}
\caption{ \label{didv}
Differential conductance as a function of dc bias 
from the non-interacting discussion. The left contact potential is oscillating.
The parameters are 
$\hbar\omega/(\gamma/2)=5$, $\epsilon^2 = 0.1$, and the Fermi 
energy is equal to the resonant energy $\mu=E_0$. For comparison 
the dashed line shows the transmission probability determined from 
the Breit-Wigner expression. 
}
\end{figure}

Photon-assisted tunneling is most easily seen either 
in the differential conductance as function of bias voltage 
$dI(V)/dV$, where
side-peaks shows up at multiples of the photon energy, 
or in the dc-current when
varying the gate potential\cite{kouwenhoven} $I(V_g)$. 
In Fig.~\ref{didv} we show an example of a $dI(V)/dV$-curve using the 
non-interacting discussion, Eq. (\ref{tiengordon}) and using 
the Breit-Wigner expression Eq. (\ref{bw}) with $U_0=0$.
The potential of the left contact oscillates. We apply a dc voltage
$V \equiv V_{1} - V_{2}$, take 
$\hbar\omega/(\gamma/2)=5$ and consider the symmetric case
$\gamma_1=\gamma_2=\gamma/2$. 
In this and all the following examples we use $\epsilon^2 = 0.1$, for which the 
expansion to second order is pertinent. 
For this choice of parameters
only the first side band peaks can be resolved. 
In a non-interacting discussion one identifies 
$U_0 = V_g$ and as a consequence both 
the differential conductance as a function of dc-voltage, 
or the current as a function of gate voltage, 
the side bands are observed at a voltage corresponding 
to the photon energy. Discussions which neglect interactions 
do not discriminate between these two methods for observing
photo-assisted transport. 
If we now consider the physically meaningful result, the theory which 
includes interactions, the general behavior will remain the same,
but the two methods of analyzing photo-assisted transport,
i. e. considering  $dI(V)/dV$ or  $I(V_g)$
now give in general different results.
The effects brought about by screening,   
discussed in more detail in the next section, are: 
First, the relative 
weight of the sidebands and the central
peaks will not be the same in the two situations. 
Second screening also 
brings about an asymmetry in the weights of the side
bands for $\pm n\hbar\omega$.
In a discussion that neglects 
interaction the side bands have the same weight.
In contrast, in the interacting case, if the equilibrium 
chemical potential does not coincide with the resonant energy,  
screening will be different for the two
voltages where peaks are seen, and accordingly their weights will differ. 
Such asymmetries are seen in experiments \cite{drexler}.
Below we discuss these effects in detail. 

\subsection{Gate driven case}

First, consider a sample subject to a dc bias $V = V_{1} -V_{2}$ 
and an oscillating voltage $V_{g}(\omega)$ applied solely at the 
gate. For simplicity we take $C_1=C_2=0$.
In this case there can be no dc photo-current when $\mu_1=\mu_2$, since
$\sum_\beta {\bf A}_{\beta\beta}(\alpha,E,E)={\bf 0}$ (see Eqs.~(\ref{photoconductance})
and (\ref{current})) as a consequence of the unitarity of the scattering matrix.
The effect of photon-assisted tunneling in this setup is controlled by the internal
potential. Thus, it is of interest to understand how it relates to the applied gate
voltage in the presence of screening.
From the selfconsistent theory 
(see Eq.~(\ref{uself})) we find for the ratio of the applied to the
external potential 
\begin{equation}
    \frac{U(\omega)}{V_{g}(\omega)} = \left[ 1+ \frac{i}{\omega C} \sum_{\alpha\beta}
	g_{\alpha\beta}^{(0)}(\omega;V) \right]^{-1}.
\label{uratio}
\end{equation}
This ratio is determined by the ac-conductances 
$g_{\alpha\beta}^{(0)}(\omega;V)$.
These ac-conductances are known.  At zero temperature, for 
the symmetric resonant tunneling barrier $\gamma_1=\gamma_2$
they are given by Fu and Dudley \cite{fu} and for the 
asymmetric case $\gamma_1\neq\gamma_2$ by 
B\"uttiker and Christen \cite{buttiker96}:

\begin{eqnarray}
    g_{11}^{(0)}(\omega) &=& g_{21}^{(0)}(\omega) \left[ \frac{\gamma_1}{\gamma_2}-\frac{\gamma}
	{\gamma_2} \left( 1-i\frac{\hbar\omega}{\gamma} \right)\right],
	\\
    g_{22}^{(0)}(\omega) &=& g_{12}^{(0)}(\omega) \left[ \frac{\gamma_2}{\gamma_1}-\frac{\gamma}
	{\gamma_1} \left( 1-i\frac{\hbar\omega}{\gamma} \right)\right],
	\\
    g_{21}^{(0)}(\omega; V) &=& g_{12}^{(0)}(\omega; -V),\\
    g_{12}^{(0)}(\omega) &=& \frac{e^2}{h} \frac{\gamma_1\gamma_2}{\gamma\hbar\omega}
	\frac{1}{1-i\frac{\hbar\omega}{\gamma}} \times \\
	&& \hspace*{-0.8cm} \times \left[
	\frac{i}{2} \ln \frac{[(\mu-\hbar\omega-E_0-W-eV/2)^2+(\gamma/2)^2]} 
	{[(\mu-E_0-W-eV/2)^2+(\gamma/2)^2]}\right. \nonumber \\
	&& \hspace*{-0.8cm} 
	+ \frac{i}{2} \ln \frac{[(\mu+\hbar\omega-E_0-W-eV/2)^2+(\gamma/2)^2]}
		{[(\mu-E_0-W-eV/2)^2+(\gamma/2)^2]} \nonumber \\
	&& \hspace*{-0.8cm}
	+ \arctan\left(\frac{\mu+\hbar\omega-E_0-W-eV/2}{\gamma/2}\right) \nonumber \\
	&& \hspace*{-0.8cm}
	- \left. 
	\arctan\left(\frac{\mu-\hbar\omega-E_0-W-eV/2}{\gamma/2}\right) \right] \nonumber.
\end{eqnarray}
%In the limit of small capacitances, the dynamic
%renormalization of the static potential is important.
%In this case, the $\mu-E_0-eV/2$ has to be replaced by 
%$\mu-E_0-eU_0-eV/2$. 

With these expressions Eq.~(\ref{uratio}), the ratio of internal to
external potential, can be evaluated.
This ratio has two simple limits. 
In the non-interacting limit $C\rightarrow\infty$, 
the internal potential directly 
follows the applied potential.
In the limit $C\rightarrow0$, 
we have a charge neutral sample and $U(\omega)=0$.

\begin{figure}
\narrowtext
\epsfysize=9cm
\epsfxsize=7cm
\centerline{\epsffile{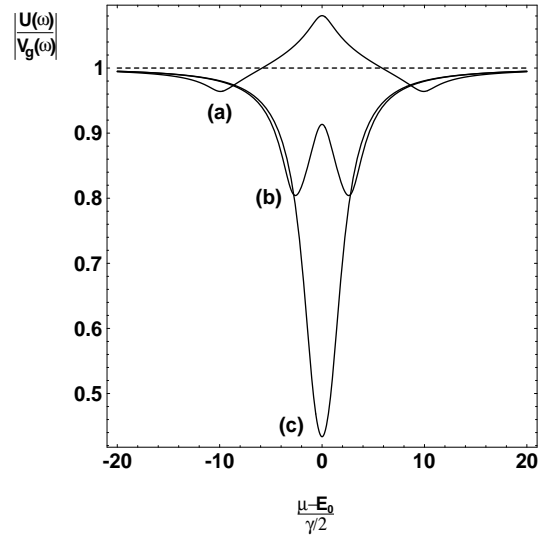}}
\caption{ \label{c1}
Ratio of the internal potential to the gate voltage as function of the 
Fermi energy, for $C=\frac{e^2}{\pi\gamma}$, $V=0$ and for the frequencies 
$a)\hspace{0.15cm} \hbar\omega/(\gamma/2)=10$, 
$b)\hspace{0.15cm} \hbar\omega/(\gamma/2)=3$, and 
$c)\hspace{0.15cm} \hbar\omega/(\gamma/2)=1$.}
\end{figure}

In Fig.~\ref{c1},
we show the absolute square ratio of the internal to
the external potential for different frequencies, 
when sweeping
the Fermi level through the resonance. The non-screened case $C\rightarrow\infty$, 
where the ratio
is 1, is shown as the dashed line. It is evident that screening introduces 
a large renormalisation of the internal
potential for this choice of 
capacitance  with a strong dependence on frequency.
One observes the largest
effect when the Fermi energy is close to the resonance. This is expected since the density in the
barrier is a Lorentzian with a peak
at resonance \cite{buttiker88}, thus providing more screening electrons.
As a function of frequency the ratio 
changes qualitatively; for low
frequencies the internal potential is reduced 
compared to the external potential, whereas
with increasing frequency the situation reverses.

\begin{figure}
\narrowtext
\epsfysize=9cm
\epsfxsize=7cm
\centerline{\epsffile{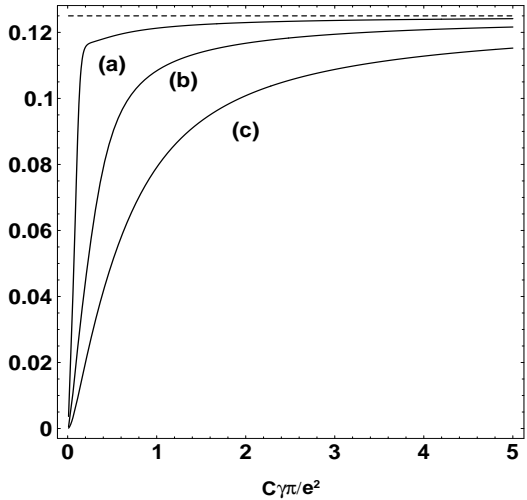}}
\caption{ \label{ratio2}
Ratio of the sideband weight to central peak weight as function of capacitance 
in the current versus gate voltage characteristic $I(V_g)$
for frequencies
$a)\hspace{0.15cm} \hbar\omega/(\gamma/2)=3$, 
$b)\hspace{0.15cm} \hbar\omega/(\gamma/2)=5$, and
$c)\hspace{0.15cm} \hbar\omega/(\gamma/2)=10$, when $\epsilon^2=0.1$.
The dashed line shows the result when no screening is present.
}
\end{figure}

Next consider a the current as a function of gate voltage. 
Since screening depends on the position
of the resonant level compared to the equilibrium electro-chemical 
potential, the central peak and the sideband will experience a different
degree of screening and, thus, their weights will no longer be given by a
Bessel function behaviour as in the non-interacting theory. 
In Fig.~\ref{ratio2}
the ratio of the sideband peak to the central peak is shown. 
The non-interacting
theory predicts a ratio of 0.125 for the parameters chosen (dashed line). 
It is seen
that, depending on capacitance and frequency, 
this ratio can be quite different.

\begin{figure}
\narrowtext
\epsfysize=9cm
\epsfxsize=7cm
\centerline{\epsffile{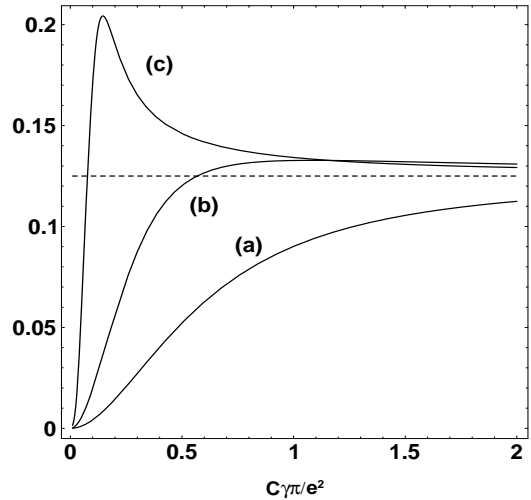}}
\caption{ \label{peakratio}
Ratio of the sideband weight to the central peak weight as function of 
the the capacitance in the differential conductance $dI/dV$ for the frequencies
$a)\hspace{0.15cm} \hbar\omega/(\gamma/2)=3$, 
$b)\hspace{0.15cm} \hbar\omega/(\gamma/2)=5$, and
$c)\hspace{0.15cm} \hbar\omega/(\gamma/2)=10$ and $\epsilon^2 = 0.1$.
The dashed line shows the result when no screening is present.
}
\end{figure}

Similarly, when measuring the differential conductance as function of dc voltage
screening will also vary as function of voltage. In this case the sideband weight 
to central peak weight ratio is shown in Fig.~\ref{peakratio}. Again, large differences
with respect to the non-interacting case is possible.

An interesting effect due to the dependence of screening on the dc voltage
(or the gate voltage) is
that sidebands will no longer be strictly Lorentzian, but skewed.
However, this skewing effect is rather small and probably difficult
to resolve experimentally.

\begin{figure}
\narrowtext
\epsfysize=9cm
\epsfxsize=7cm
\centerline{\epsffile{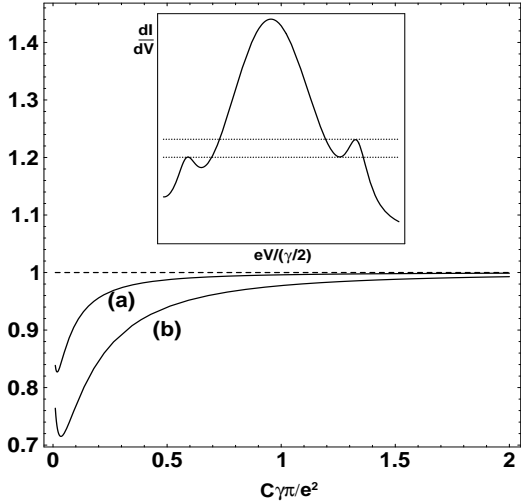}}
\caption{ \label{asymmetry}
Weight asymmetry for the $\pm\hbar\omega$ sidebands as function of capacitance
in the $dI/dV$-characteristic
for $\gamma_1/(\gamma/2)=1/4$ and $\gamma_2/(\gamma/2)=3/4$, and for 
$a)\hspace{0.15cm} \hbar\omega/(\gamma/2)=3$, $\mu/(\gamma/2)=-5$, and
$b)\hspace{0.15cm} \hbar\omega/(\gamma/2)=5$, $\mu/(\gamma/2)=-10$.
The dashed line shows the result when no screening is present. 
The insert shows an example of a differential conductance curve as a function of 
the bias voltage for $\gamma_1=\gamma_2$, $\hbar\omega/(\gamma/2) = 5 $, 
$\mu/(\gamma/2)=-10$, and $C = 0.1\frac{e^2}{\pi\gamma}$. 
The dashed line is the result without screening.
}
\end{figure}

When the Fermi level is off resonance the first sidebands corresponding to
absorbing and afterwards emitting a photon and visa versa occur at different
potentials. Screening will therefore occur asymmetrically for the two peaks
introducing an asymmetry between the $\pm$ sidebands. This effect is 
illustrated in Fig.~\ref{asymmetry}. Experimental observation of this effect
has already been made \cite{drexler}, although it has not been studied
systematically.

Another effect is visible in the inset in Fig.~\ref{asymmetry}. One notices that
the width of the central peak is significantly larger than the width of the sidebands.
Since the capacitance in this example is rather small, the charging energy is large, and
when increasing the dc bias voltage the added charge gives rise to a huge increase in the
static internal potential. The result is that the resonance floats upwards in energy,
widening the peak. For the same reason, the distance from the central peak to the sideband is
no longer simply $\hbar\omega$, but substantially larger.

\subsection{Contact driven case}

A setup often used experimentally is to couple the oscillating field to the
conductor via a bowtie antenna\cite{drexler}. In this case we assume that there
is no gate, $C_g=0$. For simplicity we take the capacitances across each
tunneling barrier to be identical, $C_1=C_2=C/2$.
The dc-current into contact 1 is then
\begin{eqnarray}
    I_1(0;V) &=& \left[ g_{111}^{(0)}(\omega,-\omega;V) \left| 
	\frac{\sum_\alpha g_{\alpha2}^{(0)}(\omega;V) + i\omega C/2}
	{\sum_{\alpha\beta} g_{\alpha\beta}^{(0)}(\omega;V)-i\omega C} \right|^2
	\right. \nonumber \\
	&& \left. + g_{122}^{(0)}(\omega,-\omega;V) \left| 
	\frac{\sum_\alpha g_{\alpha1}^{(0)}(\omega;V) +i\omega C/2}
	{\sum_{\alpha\beta} g_{\alpha\beta}^{(0)}(\omega;V)-i\omega C} \right|^2
	\right] \times \nonumber \\
	&& |V_2(\omega)-V_1(\omega)|^2.
\end{eqnarray}
For the symmetric case $\gamma_1=\gamma_2$ the dc current will vanish for zero bias,
because of the symmetry of the problem. However, in contrast to the gate driven case
a zero-bias current can be generated  for the asymmetric sample, given by
\begin{eqnarray}
\label{assym}
    I_1(0;V=0) &=& g_{111}^{(0)}(\omega,-\omega) |V_2(\omega)-V_1(\omega)|^2
	\frac{\gamma_2-\gamma_1}{\gamma} \nonumber \\
	&& \frac{ \left|\frac{g_{12}^{(0)}(\omega)}{\gamma_1\gamma_2} \right|^2
	+ \omega C \frac{Re\{g_{12}^{(0)}(\omega)\}}{\gamma_1\gamma_2}}
	{\left| \frac{g_{12}^{(0)}(\omega)}{\gamma_1\gamma_2} - 
	\frac{\omega C}{\hbar\gamma}\right|^2}.
\end{eqnarray}
Since $g_{12}^{(0)}(\omega)/(\gamma_1\gamma_2)$ is a function only of $\gamma$ we find that
the zero bias current is proportional to the effective asymmetry of the double barrier,
$\frac{\gamma_1-\gamma_2}{\gamma}$.
For small capacitances, $C\gamma_1\gamma_2/(2\hbar g_{12}^{(0)}(\omega))\ll \gamma_1,\gamma_2$,
the current is directly proportional to the asymmetry of the barrier without any
renormalisation from screening
$I_1(0;V=0)=g_{111}(\omega,-\omega)
|V_2(\omega)-V_1(\omega)|^2\frac{\gamma_2-\gamma_1}{\gamma}$.
Thus, the ac-field effectively pumps electrons through the system. 
The noninteracting Tien and Gordon discussion, 
as given by Eq.~(\ref{tiengordon}), also predicts a current
at zero bias. The result, however, being independent of the asymmetry of the system. This
prediction of a zero-bias current for the symmetric case is due to
the violations of gauge-invariance.
That a symmetric structure, in the absence of dc-voltages, cannot exhibit
a photo-current, can be understood from the following symmetry and invariance 
conditions. 
Consider first a variation of the voltage at the 
left contact $V_{1}(\omega) = V_0 \cos(\omega t)$
and suppose this produces a dc-photo current $I_{1}$.
Then consider a voltage variation of the right contact 
$V_{2}(\omega) =  - V_0 \cos(\omega t) = V_0 \cos(\omega t+ \pi)$.
By symmetry this must give a current $I_{2} = - I_{1}$. 
In reality, however, due to gauge invariance these two voltage 
oscillations are experimentally the same and hence must give 
rise to the same dc-current. But the only dc-current which reflects this 
symmetry is $I = 0$.  Clearly, the correct answer 
is a consequence of gauge invariance.

\section{Conclusion}

We have extended the scattering matrix approach to transport in phase-coherent conductors 
to take into account oscillating contact potentials and internal 
potentials in nonlinear order. The effect of 
screening has been taken into account to second order in the oscillating potentials by
means of an RPA treatment. The result is a theory, valid for arbitrary dc voltages, 
which is current and charge conserving (gauge invariant). The internal potential in
the conductor has been treated as a single parameter. Certainly, to go beyond this approximation
and treat a more realistic continuous potential distribution would 
be interesting. But even for the case
of linear ac-transport, a scattering matrix for continuous potentials exists
only to linear order in frequency\cite{buttiker932}, 
and exceptionally to second order \cite{buttikermathphys}.
Discussions of the dynamic conductance of a ballistic wire over a wide 
range of frequencies, taking into account spatial potential variations\cite{blanter98},
are not yet formulated within the scattering approach. 
It would also be interesting to extend our discussion to higher order in the 
applied voltages. For large field strengths it is possible to make one of the Bessel functions
zero, giving rising to dynamic localization\cite{holt,keay,wagner}. 
Since we find that Bessel functions can in general not give a gauge invariant answer 
it is clear that the criteria for dynamic localization will be changed in an essential 
way in the presence of interactions. 

We have applied our theory to photon-assisted tunneling using a
resonant tunneling barrier as an example. The two standard setups for photon-assisted
tunneling, applying the modulation to one of the contacts in a two-terminal experiment,
or coupling the potential to the conductor via a gate was examined within the self-consistent theory.
In both cases, the inclusion of screening leads to a renormalisation of the non-interacting answer. 
The driving field is not the applied field but the total field. Since the effective field is dependent
on screening and therefore on applied bias, chemical potential etc., the weights of the
central peak and the side peaks in the differential conduction versus 
applied voltage differ from the non-interacting
theory. Furthermore, the peak weight is no longer distributed according the increasing order of
Bessel functions. This leads to the peak ratios being a complicated function of the
screening properties of the system, and predicts an asymmetry between the corresponding
left and right sidebands. Asymmetric photoconductance peaks 
have been observed \cite{drexler,kouwenhoven}.

The necessity to include screening in the treatment of photo-assisted transport
is most clearly exemplified by the following consideration.
For a spatially symmetric system a non-interacting theory (Tien and Gordon) 
predicts a photo-current in response to the oscillation of either the 
left or the right contact voltage. In contrast, the gauge invariant 
discussion presented here, predicts that a symmetrical system exhibits no
photo-current. Our result for the two-terminal resonant tunneling barrier,
Eq.~(\ref{assym}) is a photo-current which is proportional to the asymmetry of the tunneling 
rates of the resonant double barrier structure.

In this work we have emphasized that interaction effects are important whenever
a variation of a parameter, an oscillation of a voltage, changes the charge away from 
its equilibrium value. In photo-assisted tunneling it is not sufficient 
to consider just the dc-current, but a theoretical discussion has to be self-consistent
at all frequencies. Thus there is necessarily a relation between the photo-assisted dc-current
and the displacement current. Only if the charge is investigated at all frequencies
can an electrically meaningful, that is gauge invariant, answer be found.

\acknowledgments

We are greatful for valuable discussion with Harry Thomas and Anna Pr\^etre, who
helped to clarify the derivation presented in section II.

This work was supported by the Swiss National Science
Foundation.

\appendix

\section{Current noise}

The analysis of this paper concentrates on the average zero-frequency photo-current. 
However, the approach used here also allows to find the fluctuations of the 
current. Of particular interest are the current-current correlations 
which determine the spectral densities of the current fluctuations. 
Here we present the general result 
for the noise spectra of a multi-terminal conductor in the presence of oscillating
contact potentials assuming that the internal potential is kept fixed. 
As with the average dc current a physically meaningful result requires 
in general a discussion of the effects of screening. 

For a multi-probe conductor with potentials $V_{\alpha} \cos(\omega t)$ at frequency $\omega$ 
applied to the contacts, using Eq.~(\ref{currentop}), we find the 
correlation function
\end{multicols}
\widetext
\vspace*{-0.2truein} \noindent \hrulefill \hspace*{3.6truein}
\begin{eqnarray}
    \langle \{ \Delta \hat{I}_\alpha(t+\tau), \Delta \hat{I}_\beta(t) \} \rangle &=&
	\left(\frac{e}{h}\right)^2 \int dE dE' 
	\sum_{\gamma\delta,lkl'k'} J_l\left(\frac{eV_\gamma}{\hbar\omega}\right)
	J_k\left(\frac{eV_\delta}{\hbar\omega}\right) 
	J_{l'}\left(\frac{eV_\delta}{\hbar\omega}\right)
	J_{k'}\left(\frac{eV_\gamma}{\hbar\omega}\right)
	e^{i\frac{E-E'}{\hbar}\tau} e^{i(l+l'-k-k')\omega t} \nonumber \\
	&& \mbox{Tr} [ {\bf A}_{\gamma\delta}(\alpha,E,E') 
	{\bf A}_{\delta\gamma}(\beta,E'+(l'-k)\hbar\omega,E+(k'-l)\hbar\omega) ]
	\nonumber \\
	&& [ f_\gamma(E-l\hbar\omega)(1-f_\delta(E'-k\hbar\omega))
	+ f_\delta(E'-k\hbar\omega)(1-f_\gamma(E-l\hbar\omega))]. \label{St}
\end{eqnarray}
\vspace*{-0.2truein} \hspace*{3.6truein} \noindent \hrulefill \vspace*{0.2truein}
\begin{multicols}{2}
Here the brackets $\{,\}$ denote the anti-commutator. 
In the presence of ac voltages the current-correlation function is not only a
function of the relative time $\tau$ but depends also on the absolute time $t$.
Experimentally what is of interest is the noise spectrum on a time scale long
compared to $2\pi/\omega$. Therefore we define the noise spectrum as an average
\begin{equation}
    S_{\alpha\beta}(\tau) = \frac{1}{2T} \int_0^T dt\, 
	\langle \{\Delta \hat{I}_\alpha(t+\tau), \Delta \hat{I}_\beta(t) \} \rangle,
\end{equation}
where $T=2\pi/\omega$ is the period. 
The factor $1/2$ arises because we have symmetrized the correlation function.
The spectral density is related to the current-current correlation function via 
$2\pi S_{\alpha\beta}(\Omega;\omega) \delta (\Omega+ \Omega') = (1/2)
\langle \{ \Delta \hat{I}_\alpha(\Omega) , 
\Delta \hat{I}_\beta(\Omega') \} \rangle$, 
which is just the Fourier transform of $S(\tau)$. We find
\end{multicols}
\widetext
\vspace*{-0.2truein} \noindent \hrulefill \hspace*{3.6truein}
\begin{eqnarray}
    S_{\alpha\beta}(\Omega;\omega) &=& \left(\frac{e}{\hbar}\right)^2 \int dE
	\sum_{\gamma\delta,lkk'}
	J_l\left(\frac{eV_\gamma}{\hbar\omega}\right) 
	J_k\left(\frac{eV_\delta}{\hbar\omega}\right) 
	J_{k'+k-l}\left(\frac{eV_\delta}{\hbar\omega}\right) 
	J_{k'}\left(\frac{eV_\gamma}{\hbar\omega}\right) \nonumber \\
	&& \mbox{Tr} [ {\bf A}_{\gamma\delta}(\alpha,E,E+\hbar\Omega) 
	{\bf A}_{\delta\gamma}(\beta,E+\hbar\Omega+(k'-l)\hbar\omega,E+(k'-l)\hbar\omega) ]
	\nonumber \\
	&& [ f_\gamma(E-l\hbar\omega)(1-f_\delta(E+\hbar\Omega-k\hbar\omega))
	+ f_\delta(E+\hbar\Omega-k\hbar\omega)(1-f_\gamma(E-l\hbar\omega))]. \label{S}	
\end{eqnarray}
%\vspace*{-0.2truein} \hspace*{3.6truein} \noindent \hrulefill \vspace*{0.2truein}
%\begin{multicols}{2}
In the limit of vanishing driving frequency, $\omega = 0$, Eq.~(\ref{S}) reduces to 
the frequency-dependent noise spectra of Ref.~\cite{buttiker92}.

For the special case that the scattering matrices can be taken to be independent 
of energy, i.e.\ ${\bf A}_{\gamma\delta}(\alpha,E,E+\hbar\omega)=
{\bf A}_{\gamma\delta}(\alpha)$ Eq.~(\ref{S}) simplifies considerably. 
Using the addition theorem for Bessel functions we find 
%\end{multicols}
%\widetext
%\vspace*{-0.2truein} \noindent \hrulefill \hspace*{3.6truein}
\begin{eqnarray}
    S_{\alpha\beta}(0;\omega) &=& \left( \frac{e}{\hbar} \right )^2
	\int dE \sum_{\gamma\delta,l} \mbox{Tr} [ {\bf A}_{\gamma\delta}(\alpha)
	{\bf A}_{\delta\gamma}(\beta)] 
	J_l^2\left(\frac{e(V_\delta-V_\gamma)}{\hbar\omega}\right) \times \nonumber \\
	&& [ f_\gamma(E+l \hbar\omega)(1-f_\delta(E))+ 
	f_\delta(E)(1-f_\gamma(E+l \hbar\omega))]. \label{S0} 
\end{eqnarray}
\vspace*{-0.2truein} \hspace*{3.6truein} \noindent \hrulefill \vspace*{0.2truein}
\begin{multicols}{2}
For a two-terminal conductor this result is identical to that of 
Lesovik and Levitov \cite{lesovik} even though in that work this result was derived
in response to an electric field and not as here as a response to an 
oscillating contact voltage. In the experiment of Schoelkopf et al.\cite{schoelkopf} the shot
noise is measured in the presence of an oscillating voltage applied to the 
contacts of the sample.  

We emphasize that the noise spectra given by Eqs.~(\ref{S}) and (\ref{S0}) 
give only the noise for fixed internal potential. 
We have already remarked 
that the average dc current exhibits an external response 
due to photo-assisted transport only if the transmission probabilities 
exhibit an energy dependence (see Eq.~(\ref{photoconductance1})). 
In contrast, in the shot-noise spectra, we have an effect even if the scattering matrix is taken 
to be energy independent. That is a consequence of the fact that the noise spectra 
depend in a non-linear way on the Fermi functions.

\end{multicols}

\end{document}